# Capital Structure Adjustment Speed and Expected Returns: Examination of Information Asymmetry as a Moderating Role

Masoud Taherinia[1*], Mehrdad Matin[2], Jamal Valipour[3], Kavian Abdolahi[4], Peyman Shouryabi[5], Mohammad Mahdi Barzegar[6]

[1*] Department of Accounting, Faculty of Management and Economics, Lorestan University, Khorramabad, Iran. Email: taherinia.m@lu.ac.ir (*Corresponding Author*).
[2] Department of Economy & Accounting, Islamic Azad University, South of Tehran Branch, Tehran, Iran.
[3] Department of management, Financial Management, Azad University, Tehran, Iran.
[4] Department of Finance, Khatam University, Tehran, Iran.
[5] University of Rome Tor Vergata, Italy.
[6] Department of Finance, Sheldon B. Lubar College of Business, University of Wisconsin Milwaukee, WI, USA

*Abstract:* Shareholders' expectations of stock returns and fluctuations are constantly changing due to restrictions in financial status and undesirable capital structure, which constrain managers to limit the changes in price trends in order to cover the risk instigated and infused by the unfavorable situation. The present research examines the moderating impact of information asymmetry on the relationship between capital structure adjustment and expected returns. The data from 120 companies approved in the Tehran Stock Exchange were extracted, and a hybrid data regression model was used to test the research hypotheses. Findings indicate that the capital structure adjustment speed correlates with the expected returns. Moreover, the information asymmetry positively affects the relationship between capital structure adjustment speed and expected returns.

## Introduction

The flow of information in the market environment touches on the behavior of market participants. Environmental change creates the conditions for market participants to have a different share of this information flow (Bushman & Smith, 2001). Empirically, people also have different information (Hessari et al., 2023; Mundi, 2021). The information affects their behavior often and indicates information asymmetry between the two parties in a deal. This information asymmetry is due to the different flow of information among market participants (Lambrecht & Myers, 2012). Companies have exclusive ownership of their information in a way that prevents them from transferring their added value to investors. Attempting to obtain information through other channels in the capital market leads to the imposition of unrealistic information risk on the investor, and a part of the expected return is limited to the cost of accessing information. In connection with accounting information, there are two main characteristics: the quality of this information and its distribution (Kurdistani & Rahimkhani

M. (2016). Accounting information risk can be divided into two components in the context of the capital market environment: the part related to the ambiguity on the accuracy of this information and the part related to the distribution of this information. Ambiguous accounting information undermines the relationship between accounting figures and economic facts, and consequently, it increases information asymmetry. Therefore, having a favorable accounting information environment increases the ability of financial reports to transfer company information (Patatoukas & Thomas, 2016; Etemadi et al., 2022; Moezzi et al., 2012). On the other hand, it causes the equal distribution of this information among market participants (Vesal et al., 2013; Takalo et al., 2013; Dehghanan et al., 2021; Gogheri et al., 2013). Assuming the stickiness of investments, the capital structure must absorb the shocks to the company. The ability of the company to adjust the capital structure leads to the provision of smoothing conditions for the division of profits. In other words, information asymmetry is expected to decrease as the rate of capital structure adjustment accelerates. The inability of companies to adapt to environmental conditions due to fluctuations leads to an increased risk of poor selection for investors, and consequently, capital costs and changes in investors' expectations will be formed. The created risk can be partially covered by identifying the trend and stability of the company based on the capital structure adjustment speed (Martínez et al., 2018; Hashemzadeh et al., 2011; Gheitarani et al., 2022a; Ghazinoory et al., 2017; Hessari & Nategh, 2022a). In previously conducted research, those of information asymmetry and information distribution with the information transparency approach have been examined, and the direct impact of information asymmetry on information quality has been highlighted. Furthermore, the quality of information regarding financial statements and capital markets has been examined. In this study, we intend to explain the effect of capital structure adjustment on changes in shareholder expectations by separating information asymmetry into two parts: data collection and pricing the investment decisions.

Expected returns are essential for investment and corporate financing decisions, and it is inherently important for financial professionals to pay attention to the expected returns. On the other hand, one of the salient features of inefficient markets is the significant information asymmetry between entrepreneurs and investors. Investors in these markets should rely heavily on the contents of financial statements while making investment decisions, and since processes are designed to estimate expected returns, historical data have been used to calculate important input factors, and therefore, investment risk will increase. In examining the role of disclosure quality on stock pricing efficiency from both quantitative and qualitative dimensions, it was concluded that the price discovery process is improved, and stock pricing becomes closer to reality by improved quality and quantitative disclosure.

Information asymmetry is the knowledge or information gap (Bai et al., 2024) between investors. This refers to a trading situation in which some members have more information about the market situation than others and have different skills and backgrounds in information processing (Gheitarani et al., 2022b; Hakkak et al., 2016; Hessari & Nategh, 2022b; Nawaser et al., 2015). As a result, increasing information asymmetry enhances the net cost of capital. Consequently, the expected return of investors increases Denis & McKeon (2012). Financial players need to gather information and have the skills and intelligence to analyze them in favor of a better financial position. This relates to the well-studied notion of a different kind of intelligence, which Bai et al. (2023) studied in a business setting. This kind of intelligence can be useful when studying how and why investors make their financial decisions regarding the stock market. According to researchers, this "intelligence contributes significantly to emotional resilience" (Bai et al., 2023), which is of critical importance in the psychology of trade for investors and traders of the stock market.

In a situation where companies have the ability to finance outside the organization but with high financing costs, they adjust their leverage ratios.[9] However, modification costs prevent rapid correction and achievement of the objective function. The company pays adjustment costs in order to avoid the costs of a non-optimal capital structure.[10] However, the capital structure of financial institutions is determined by asymmetric information as a determinant of cash flows and stock prices. Jensen and Meckling (1976) showed that increasing the level of debt leads to agency costs. Further debt leads to a loss of opportunities to improve the value of the company and increases the costs of supervision and management (Charles et al., 2017; Dehkordy et la., 2013; Shamsaddini et al., 2015). Miller (1997) argues that changes in capital structure due to debt growth led to problems with underinvestment. On the other hand, information asymmetry occurs when companies issue their financial resources in the form of equity (issue of shares). At the time of publication of this share, investors would think that managers have more information about the company and its value, and therefore, managers make use of this valuation (Chung et al., 2018; Jahanshahi et al., 2020; Khaksar et al., 2010; Asadollahi et al., 2011). Therefore, investors are demanding higher returns to compensate for their ignorance. This increases the cost of capital and, ultimately, the capital structure adjustment.[8] Information asymmetry, especially between management and investors, leads to the problem of unfavorable selection in corporate financing, increases the sensitivities of the financing process, and can lead to the deviation of the capital structure from the optimal level (Gong et al., 2021). Unfavorable adjustment in capital structure leads to changes in investment decisions and to a certain level of change in information symmetry between managers and shareholders in such a way that project managers plan investment projects for their personal and opportunistic interests (Callen et al., 2010; Armstrong et al., 2011). Capital adjustment is a function of the speed of change of debt ratios, adjustment costs, and the availability of suitable investment opportunities. In a situation where the company does not have suitable investment opportunities, it has no desire to publish the relevant bad news, which leads to increased information asymmetry between the company and investors. Leary and Michael (2011) in their study, concluded that the low unused debt capacity in companies leads to reduced smoothing of dividends. These companies have avoided reducing dividends, and in response to declining profitability and to offset future adverse effects, they do not change the dividend trend. The ability to restructure capital depends on the adjustment costs that the company always faces. In this regard, companies may use short-term strategies to avoid the costs of capital restructuring. Still, these strategies only work efficiently if they do not create costs for the company (including lowering its credit rating in the capital market). Also, Faulkender (2012) argued that the dividend-paying companies are adjusting their capital structure faster than other companies. This approach originated from the company's goal of maintaining low-risk financing capacity as well as the incentive to access low-cost capital markets. In case the cost of adjusting the capital, structure is low, information asymmetry decreases due to the speed of correction and change in the capital structure.

According to the review of the literature of the conducted research, and with regards to the purpose of the current study, which is limited to the study of pricing information asymmetry based on financing constraints and capital structure adjustment, the research hypotheses are presented as follows:

**First Hypothesis:** The capital structure adjustment speed has a significant impact on the expected returns.
**Second Hypothesis:** Information asymmetry has a significant impact on the relationship between capital structure adjustment and expected returns.

## Methodology

The present research is of a quantitative type. Since the concepts under study exist objectively in the outside world, it is categorized under the positivist paradigm and empirical studies, where the reality of the subject and the researcher are completely independent of each other. The predominant methodology of such research is the use of multiple statistical methods. Since this research method is based on describing the real relationships between the available data (description of what exists), which is expressed in the form of a model, the research is also descriptive. Modeling is the process of formulating relationships in the real world. Therefore, in this study, the real aspects of relationships are first studied, and then the model is designed based on the relevant assumptions and relationships. The modeling process in this research is based on the scientific method and in the form of the theoretical logic.

This research examines the pricing of information distribution based on financing constraints and capital structure adjustment. For this purpose, the following model has been used to test the hypotheses:

$$RET_{it} = \beta_0 + \beta_1 ST_{it} + \beta_2 AS_{it} + \beta_3 AS * ST_{it} + \beta_4 A\_QU_{it} + \beta_5 F\_PE_{it} + \beta_6 CO_{it} + \beta_7 NON_{it} + \beta_8 SI_{it} + \beta_9 LO_{it} + \beta_{10} LEV_{it} + \varepsilon \quad \ldots(1)$$

Where $RET$, is expected return; $AS$, information asymmetry; $St$, capital structure adjustment; $A\_QU$, accounting information quality; $F\_PE$, future performance; $CO$, conservatism in accounting; $NON$, environmental uncertainty; $SI$, company size; $LEV$, financial leverage; and $LO$, company Losses.

Expected Return: The expected return of $i$ share over the 12-month period is based on The Fama French 3-Factor Model.

Information Asymmetry: The information asymmetry variable is calculated as follows:

$$SP_{it} = \frac{AP_{it} - BP_{it}}{P} * O\_SI \quad \ldots(2)$$

Where $BP_{i,t}$, best bid price to buy the company $i$ share at the $t$ moment; $AP_{i,t}$, best bid price to sell the company $i$ share at the $t$ moment; $O\_SI$, share order size; and P, average price of quotations.[19]

Capital Structure Adjustment: The Fliers model is used to calculate the cost of capital structure adjustment. In this model, the speed of capital structure adjustment is considered as a factor to create costs. Thus, if the speed of capital structure adjustment is high, less correction costs will be imposed on the company. In the first step, the following model is estimated for each year:

$$DR_{it+1} = \beta_0 + \beta_1 DR_{it} + \beta_2 PR_{it} + \beta_3 MTB_{it} + \beta_4 EXP_{it} + \beta_5 TA_{it} + \beta_6 TR_{it} + \beta_7 SI_{it} + \varepsilon \quad \ldots(3)$$

Where DR, is debt ratio; PR, ratio of profitability to total assets; MTB, ratio of firm's book value to its market value; EXP, ratio of investment costs to total assets; TA, total fixed assets to total assets; TR, two-year returns; and SI, logarithm of total assets.

The residual error of the above model (DR) is first moderated and normalized based on the error minimization method and then enters the following model:

$$DR_{it+1} - DR_{it} = \beta_0 + \beta_1(DR*_{it} - DR_{it}) + \varepsilon \quad \ldots(4)$$

The $\beta_1$ coefficient indicates the capital structure adjustment speed.

Company Size: Equivalent to the normal logarithm is the sum of the company's assets.
Conservatism in Accounting: In this research, the following model has been used to calculate conservatism in accounting following. In this study, Basu model (1997) has not been used due to serial alignment between the interactive effect of return, overconfidence, environmental uncertainty and funding constraints, as well as deviation in measurement.

$$RE_{it} = logRE_{it} + \log(1 + ROE)_{it} + \log\left(\frac{B}{M}\right)_{it} + \varepsilon_{it} \quad \ldots(5)$$

Where, RE, stock returns; ROE, capital return and B/M, ratio of firm's book value to its market value in the shareholders' equity.

Future Performance: The average ratio of operating cash flow extracted from the cash flow statement to the total assets at the beginning of the period has been used to evaluate the performance in this research. This ratio has been used to reduce the error caused by profit manipulation.

Environmental Uncertainty: The standard deviation of changes in sales revenue over a period of 3 years is used to measure environmental uncertainty. The use of standard deviation to measure environmental uncertainty has been proposed and used by researchers namely Dichev and Tang (2009), and Anwari Rostami and Kiani (2015).

Loss: It is an imaginary variable that if the company has a loss this year, the last year or during the last two years, it equals one, otherwise, zero.

Financial Leverage: is equal to the ratio of total liabilities to total assets of the company.

Accounting Information Quality (AIQ): The model of [Francis] et al., (2005) has been used to evaluate the quality of accounting information in this research:

$$TCA_{j,t} = \alpha_0 + \alpha_1 CFO_{j,t-1} + \alpha_2 CFO_{j,t} + \alpha_3 CFO_{j,t+1} + \alpha_4 \Delta REV_{j,t} + \alpha_5 PPE_{j,t} + \varepsilon_{j,t} \quad \ldots(6)$$

Where TCA is the total accruals of j company in t year; CFO is cash flow from corporate operations; ΔREV is a change in the company's sales revenue; and EPP is net fixed asset turnover.

The above model is estimated for each year-industry, and the reverse 3-year standard deviation, except error, is considered a variable of accounting information quality.

The statistical population of this research is the companies accepted by the Tehran Stock Exchange that have been listed on the Tehran Stock Exchange since 2011. The application of constraints limits the statistical population due to the specific time interval for testing and extracting prices. In other words, a statistical sample includes a hypothetical statistical population that is assumed to have the following constraints:

1. Their fiscal year should end on March 20. This restriction is due to the comparison capability to evaluate and estimate variables. In addition, there would be a basis for selecting the date of extraction of share prices.
2. From 2011 to 2020, their trading symbol has not been closed for more than one-third of the trading days of the year. This limitation is also due to the measurement of variables, especially that of financial flexibility. Closure of the trading symbol of a stock for a long time will cause the reluctance of shareholders and, thus, reduce the competitiveness of the stock.
3. The company should not be of the financial intermediation and leasing industries type. The reason for this is that such companies themselves are considered institutional investors active in the capital market, and the impact of their activities has been seen in the models. Moreover, these companies are subject to special mechanisms that are different from other companies active in the industry.

Eventually, the companies surveyed in this study included 120 companies accepted by and listed on the Tehran Stock Exchange.

**Results and Discussion**

To provide an overview of the important characteristics of the calculated variables, the following table shows some descriptive statistics of these variables: mean, median, standard deviation, minimum and maximum observations.

Table 1: Descriptive Statistics of Research Variables.

| Variable | Mean | Median | SD | Max. | Min. |
|---|---|---|---|---|---|
| Expected Return | 0.1709 | 0.1497 | 0.7029 | 0.4732 | 0.1011 |
| Information Asymmetry | 0.0651 | 0.0315 | 0.4317 | 1.0628 | -0.0295 |
| Capital Structure Adjustment Speed | 0.3110 | 0.0453 | 0.1252 | 0.9041 | -0.5119 |
| Accounting Information Quality | 0.4110 | 0.2110 | 0.9198 | 1.1956 | 0.0098 |
| Company Size | 5.9417 | 5.8864 | 0.6197 | 8.5200 | 4.2912 |
| Accounting Conservatism | -0.0004 | -0.0005 | 0.0035 | 0.0328 | -0.0057 |
| Future Performance | 0.0825 | 0.0063 | 0.2487 | 2.3595 | 0.0100 |
| Environmental Uncertainty | 0.1345 | 0.0965 | 0.1181 | 0.8487 | 0.0016 |
| Company Losses | 0.0406 | 0.0000 | 0.1976 | 1.0000 | 0.0000 |
| Financial Leverage | 0.6339 | 0.6441 | 0.2098 | 1.7403 | 0.0405 |

The average information asymmetry was 0.651, which indicates the limited rate of liquidity held by the under-study companies that can be justified by the expected rate of return of 17 percent. The average capital structure adjustment variable is equal to 0.31, which indicates the long process of capital structure adjustment among the under-study companies. The average accounting information quality equals 0.4110, which shows a high fluctuation in the cash flows created in recent years. The average future performance equals 0.0825, which indicates the limited rate of liquidity held by the selected companies. The average of accounting conservatism variable is -0.0004, which reveals limited conservatism among the companies. An average of 13 percent of environmental uncertainty indicates instability in the sales process of the stock exchange companies. The average of financial leverage is 0.63 which shows a major part of the company's expenses are financed from debts.

Table 2 shows the results of estimating the research model in the present study.

Table 2: Results of research model estimation

| Variable | Model 1 | | | |
|---|---|---|---|---|
| | Coefficient | t Statistic | Probability | VIF |
| Information Asymmetry (A) | -0.00280 | -2.1939 | 0.0299 | 1.15 |
| Capital Structure Adjustment (C) | 1.7712 | 2.7849 | 0.0054 | 1.44 |
| Information Asymmetry * Capital Structure Adjustment | 1.3577 | 2.6295 | 0.0085 | 1.67 |
| Accounting Information Quality | 0.0691 | 2.1078 | 0.0354 | 1.41 |
| Accounting Conservatism | -5.7361 | -1.5171 | 0.1296 | 1.73 |
| Financial Leverage | 0.0125 | 1.8860 | 0.0597 | 1.25 |
| Company Losses | -0.0567 | 0.5421 | 0.5878 | 1.91 |
| Future Performance | 0.4312 | -4.1152 | 0.0000 | 2.02 |
| Company Size | 0.1124 | 0.2639 | 0.7919 | 1.16 |
| Environmental Uncertainty | 0.0942 | -0.8836 | 0.3772 | 1.34 |
| Fixed Component | -0.3326 | -2.2504 | 0.0301 | - |
| Impacts of Year | Controlled | | | |
| Impacts of Industry | Controlled | | | |
| Determination Coefficient | 4216.0 | | | |
| Modified Determination Coefficient | 3980.0 | | | |
| Durbin-Watson (DW) Statistic | 9617.1 | | | |
| F Statistics | 1720.10 | | | |
| F Statistical probability | 0.0000 | | | |

Based on model 1, the first research hypothesis, the capital structure adjustment speed significantly impacts the expected return, is confirmed since it is less than 5 percent (t-statistic, 2.7849).

The second research hypothesis, information asymmetry, has a meaningful impact on the relationship between the capital structure adjustment speed and the expected return and has not been rejected. As can be observed and according to Table 4, the t statistic is equal to 2.6295, which indicates that the second hypothesis is confirmed at a 5 percent error level.

**Conclusions and Suggestions**

This research studied the moderating impact of information asymmetry on capital structure adjustment and expected return. The first hypothesis of the research has been confirmed. The results reveal that the capital structure adjustment speed has led to a change in expected returns due to an increase in information asymmetry, the management measures are based on short-term results, and personal interests are prioritized. The results of this hypothesis are consistent with Alti's research (2003). While the representation problems are usually due to managers' over-investment with regard to personal interests (Daneshmandi et al., 2023), and it simply leads to underinvestment in companies with limited funding, modern investment theory is of the opinion that companies are willing to invest in projects with fixed rates of return which are higher than the cost of capital and require less financing costs.[28] In case of limited funding, the quality of information provided to investors will be reduced, increasing the information gap between managers, shareholders, and investors.

The results of examining the second hypothesis of the research indicate that information asymmetry has a meaningful impact on the relationship between capital structure adjustment and expected returns. Investors have diverse abilities to process information about the company's performance. Therefore, the capital structure adjustment speed and the absorption of shocks enforced on it can lead to the creation of favorable conditions for uninformed investors in information analysis and eventually reduce the pricing of information asymmetry (earning abnormal returns) in financial markets. The results of this hypothesis are consistent with that of Baum. They proved that the process of adjusting the capital structure is asymmetric and remains a function of risk in the leverage ratio of the company and its financial condition. In a situation where the company can cope with the shocks, it controls the relevant fluctuations by adjusting the capital structure.